\def\mode{1}
\documentclass[sigconf, nonacm, screen=true, natbib=true]{acmart}

\usepackage[a-2b]{pdfx}

\newcommand\vldbdoi{10.14778/3611540.3611578}
\newcommand\vldbpages{3890 - 3893}
\newcommand\vldbvolume{16}
\newcommand\vldbissue{12}
\newcommand\vldbyear{2023}
\newcommand\vldbauthors{\authors}
\newcommand\vldbtitle{\shorttitle} 
\newcommand\vldbpagestyle{empty}

\title[{A Tutorial on Visual Representations of Relational Queries}]{A Tutorial on Visual Representations of Relational Queries}

\author{Wolfgang Gatterbauer}
\orcid{0000-0002-9614-0504}%
\affiliation{%
    \orcidicon{0000-0002-9614-0504}
    \institution{Northeastern University}	
    \city{Boston}
    \state{Massachusetts}
    \country{USA}
}
\email{w.gatterbauer@northeastern.edu}

\usepackage{url}
\usepackage{enumitem} %

\usepackage{upgreek} %

\usepackage{listings} %

\usepackage{tabularx} %
\usepackage{booktabs} %
\usepackage{multirow}
\usepackage{hhline} %
\usepackage{pgfplotstable} %

\usepackage{easybmat} %

\usepackage{subcaption} %
\usepackage{rotating}		%

\usepackage{xspace} %

\usepackage[ruled,noend,linesnumbered]{algorithm2e} %

\DontPrintSemicolon     %
\SetNlSty{}{}{}                %
\SetAlgoInsideSkip{smallskip}   %
\SetAlCapSkip{1.5mm}             %
\SetInd{1.5mm}{1.5mm}             %

\SetAlFnt{\small}			%
\SetAlCapFnt{\small}		%
\SetAlCapNameFnt{\small}

\newtheorem{definition}{Definition}

\newtheorem{questionW}{Question}
\newtheorem{resultW}{Result}

{\end{itshape}
}
\let\footnotesize\scriptsize

\newcommand{\hide}[1]{}

\usepackage{hyperref}   %
\hypersetup{                                                            %
	pdfpagemode=UseNone,               %
    pdfstartview=Fit,
    pdfpagelayout=SinglePage,       %
    bookmarksnumbered=true,
    bookmarksopen=false,             %
    unicode,
    colorlinks=true,       %
    linkcolor=blue,          %
    citecolor=cyan,  %
    filecolor=magenta,      %
}

\usepackage[capitalise,nameinlink]{cleveref} %

\crefname{algocf}{alg.}{algs.}
\Crefname{algocf}{Algorithm}{Algorithms}
\crefalias{AlgoLine}{line}%
\if 1\mode
	\AtEndPreamble{%
	    \hypersetup{colorlinks,
	      linkcolor=purple,
	      citecolor=blue,
	      urlcolor=ACMDarkBlue,
	      filecolor=ACMDarkBlue}}
\fi

\if 0\mode
	\makeatletter 

	\renewcommand{\@opargbegintheorem}[3]{%
	    \parskip 0pt %
	    \trivlist
	    \item[%
	    	\hskip 10\p@										%
	        \hskip \labelsep
	        {\sc #1\ #2\             %
	   \setbox\@tempboxa\hbox{(#3)}  %
	        \ifdim \wd\@tempboxa>\z@ %
	            \hskip 0\p@\relax    %
	            \box\@tempboxa       %
	        \fi.}%
	    ]
	    \it
	}

	\def\@begintheorem#1#2{%
	    \parskip 0pt %
	    \trivlist
	    \item[%
	    	\hskip 10\p@										%
	        \hskip \labelsep
	        {{\sc #1}\hskip 5\p@\relax#2.}%
	    ]
	    \it
	}

	\makeatother
\fi

\usepackage{tcolorbox}		%
\tcbuselibrary{breakable,skins}		%
\tcbset{
		enhanced jigsaw,	%
		colback=gray!20,
		colframe=gray!40,
		arc=0mm,
		boxrule=1pt,		%
		left=1pt,
		right=1pt,
		topsep at break=1mm,			%
		top=1pt,		%
		bottom=0mm,		%
		breakable,		%
		parbox = false		%
		}

\makeatletter
\DeclareRobustCommand*\uell{\mathpalette\@uell\relax}
\newcommand*\@uell[2]{
  \setbox0=\hbox{$#1\ell$}
  \setbox1=\hbox{\rotatebox{10}{$#1\ell$}}
  \dimen0=\wd0 \advance\dimen0 by -\wd1 \divide\dimen0 by 2
  \mathord{\lower 0.1ex \hbox{\kern\dimen0\unhbox1\kern\dimen0}}
}
\makeatother

\newcommand{\smallsection}[1]{\vspace{2mm}\noindent\textbf{#1.}} %

\renewcommand{\epsilon}{\varepsilon} %

\usepackage{booktabs} %
\graphicspath{{./figs/}}
\usepackage{numprint}
\usepackage{multirow}

\usepackage{scalerel}
\usepackage{tikz}
\usetikzlibrary{svg.path}

\definecolor{orcidlogocol}{HTML}{A6CE39}
\tikzset{
  orcidlogo/.pic={
    \fill[orcidlogocol] svg{M256,128c0,70.7-57.3,128-128,128C57.3,256,0,198.7,0,128C0,57.3,57.3,0,128,0C198.7,0,256,57.3,256,128z};
    \fill[white] svg{M86.3,186.2H70.9V79.1h15.4v48.4V186.2z}
                 svg{M108.9,79.1h41.6c39.6,0,57,28.3,57,53.6c0,27.5-21.5,53.6-56.8,53.6h-41.8V79.1z M124.3,172.4h24.5c34.9,0,42.9-26.5,42.9-39.7c0-21.5-13.7-39.7-43.7-39.7h-23.7V172.4z}
                 svg{M88.7,56.8c0,5.5-4.5,10.1-10.1,10.1c-5.6,0-10.1-4.6-10.1-10.1c0-5.6,4.5-10.1,10.1-10.1C84.2,46.7,88.7,51.3,88.7,56.8z};
  }
}

\RequirePackage{etoolbox}
\DeclareRobustCommand\orcidicon[1]{\href{https://orcid.org/#1}{\mbox{\scalerel*{
\begin{tikzpicture}[yscale=-1,transform shape]
\pic{orcidlogo};
\end{tikzpicture}
}{|}}}}

\usepackage{xspace}            %

\newcommand{\queryvis}{\textsf{QueryVis}\xspace}
\newcommand{\diagrams}{\textsf{Relational Diagrams}\xspace}

\begin{document}

\begin{abstract}

Query formulation is increasingly performed by systems that need to guess a user's intent 
(e.g.\ via spoken word interfaces). 
But how can a user know that the computational agent is returning answers to the {``right''} query? 
More generally, 
given that relational queries can become pretty complicated, 
\emph{how can we help users understand existing relational queries}, whether human-generated or automatically generated?
Now seems the right moment to revisit a topic that predates the birth of the relational model:
developing visual metaphors that help users understand relational queries.

This lecture-style tutorial 
surveys the key \emph{visual metaphors developed 
for visual representations of relational expressions}. 
We will survey the history and state-of-the art of relationally-complete 
diagrammatic representations of relational queries,
discuss the key visual metaphors developed in over a century of 
investigating diagrammatic languages,
and organize the landscape by mapping their used visual alphabets to the syntax and semantics of Relational Algebra (RA) and Relational Calculus (RC).
\end{abstract}

\maketitle

\pagestyle{\vldbpagestyle}
\begingroup\small\noindent\raggedright\textbf{PVLDB Reference Format:}\\
\vldbauthors. \vldbtitle. PVLDB, \vldbvolume(\vldbissue): \vldbpages, \vldbyear.\\
\href{https://doi.org/\vldbdoi}{doi:\vldbdoi}
\endgroup
\begingroup
\renewcommand\thefootnote{}\footnote{\noindent
This work is licensed under the Creative Commons BY-NC-ND 4.0 International License. Visit \url{https://creativecommons.org/licenses/by-nc-nd/4.0/} to view a copy of this license. For any use beyond those covered by this license, obtain permission by emailing \href{mailto:info@vldb.org}{info@vldb.org}. Copyright is held by the owner/author(s). Publication rights licensed to the VLDB Endowment. \\
\raggedright Proceedings of the VLDB Endowment, Vol. \vldbvolume, No. \vldbissue\ %
ISSN 2150-8097. \\
\href{https://doi.org/\vldbdoi}{doi:\vldbdoi} \\
}\addtocounter{footnote}{-1}\endgroup

\section{Introduction}

\begin{figure}[tb]
\centering
\includegraphics[scale=0.31]{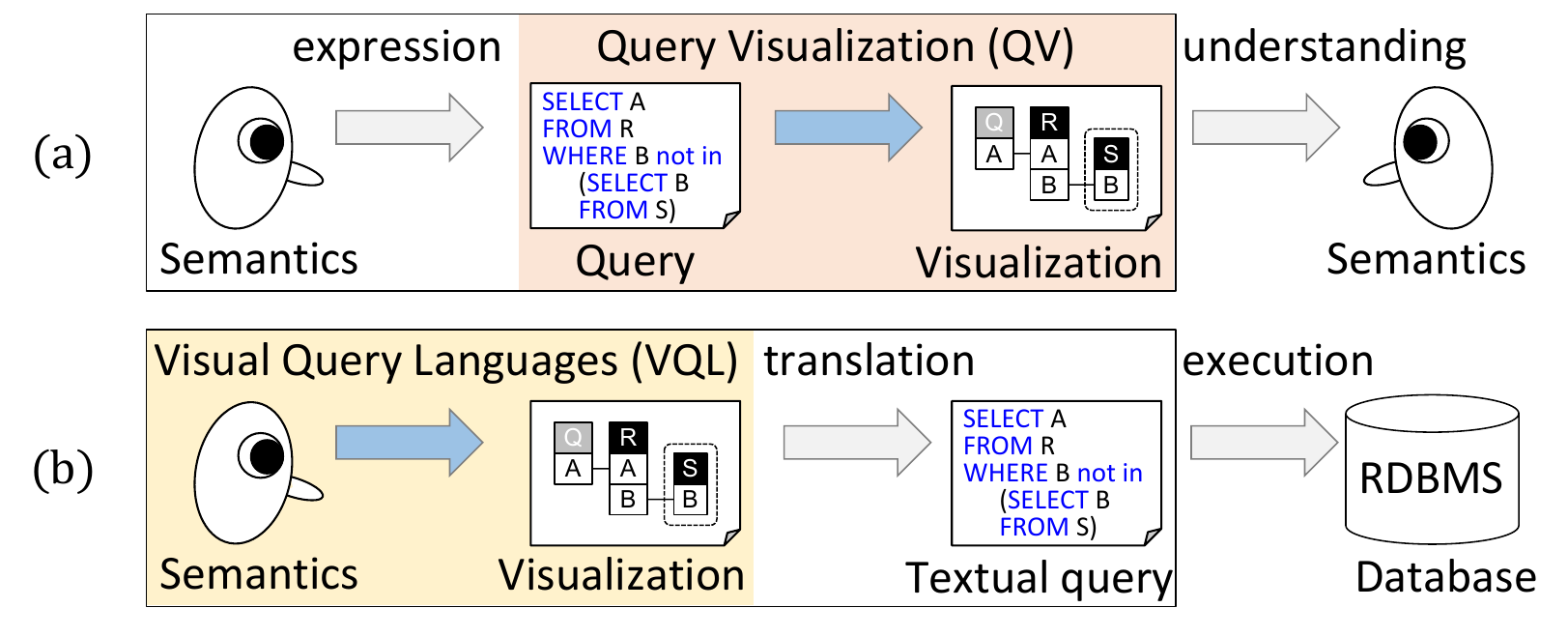}
\caption{Contrasting 
Query visualization (on the top (a), in orange)
with
Visual Query Languages (VQL) (on the bottom (b) in yellow).
}
\label{Fig_QueryVisualization}
\end{figure}

The design of relational query languages and the difficulty for users to compose relational queries have
received much attention over the 
last 40 years~\cite{DBLP:journals/vlc/CatarciCLB97, 
ChanUserDatabaseInterface:1993,
Harel:Nonprocedural:1985,
FrameworkForChoosingQueryLanguages:1985,
LEGGETT1984493,
DBLP:journals/csur/Reisner81, 
Reisner1975:HumanFactors,
Welty-Stemple:1981,
scamell:1993}.
A complementary and much-less-studied problem is that of helping users 
\emph{read and understand an existing relational query}.  
Reading code is hard, and SQL is no exception.  
With the proliferation of public data sources, and associated queries, users increasingly have a need to read other people's queries and scripts.  
Furthermore, it is usually much easier to modify a draft than to write something from scratch.  
As such, modifying an already existing query could be an effective way to write new queries.
However, modifying an existing query requires first to understand it. 
For that reason, it is valuable to help users understand queries,
and visualization is one obvious route.
While visual methods for expressing queries have been studied
extensively in the database literature under the topic of Visual Query Languages (VQLs)~\cite{DBLP:journals/vlc/CatarciCLB97}, 
the challenges for supporting the explicit \emph{reverse functionality} of creating a visual representation of an existing query
(``Query Visualization'') are on a whole different from the problem of composing a new query (\cref{Fig_QueryVisualization}).

\begin{figure}[tb]
\includegraphics[scale=0.31]{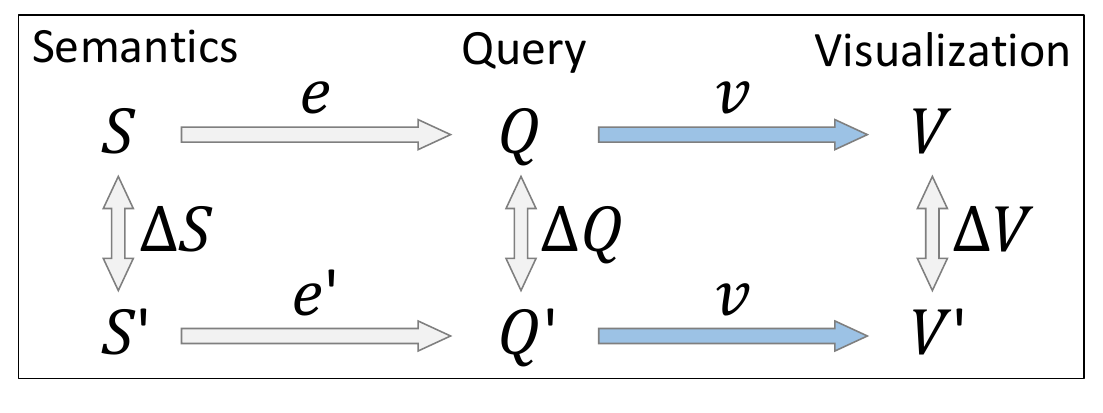}
\hspace{12mm}
\caption{A simple algebraic framework will help us discuss various principles of query visualization
(best understood with \hyperref[Fig_QueryVisualization]{Fig.~\ref{Fig_QueryVisualization}a}).}
\label{Fig_QueryVisualizationAlgebraic}
\end{figure}

The tutorial uses a few relational queries 
to survey and summarize over a history of \emph{diagrammatic (thus visual) representations 
of first-order logic queries and statements}.
The tutorial will contrast and highlight similarities and differences
between approaches proposed across communities
and use a mapping of the visual representations to equivalent expressions in 
Relational Algebra (RA) and Relational Calculus (RC) to guide the journey.

\smallsection{Outline}
The lecture-style 1.5-hour tutorial consists of five parts:

(1) \emph{Why visualizing queries and why now}:
	We contrast Query Visualization (QV) with Visual Query Languages (VQL) 
	and give several usage scenarios 
	for the use of the former.

(2) \emph{Principles of Query Visualization}:
	We discuss 8 recently proposed principles of query visualization~\cite{GatterbauerDJR:PrinciplesOfQV:2022,gatterbauer2011databases},
	re-phrased in the terminology of ``Algebraic Visualization Design''~\cite{DBLP:journals/tvcg/KindlmannS14} (\cref{Fig_QueryVisualizationAlgebraic}).
	We later refer to them when discussing different visualizations.

(3) \emph{Logical foundations of relational query languages}:
	We discuss the logical foundations of relational query languages.
	These concepts are also used 
	when discussing visual representations.

(4) \emph{Early diagrammatic representations}:
	Diagrammatic representations for logical statements
	were developed well before relational databases. 
	We will discuss the influential beta existential graphs by Peirce~\cite{peirce:1933}
	and their connection to the much later developed RC.

(5) \emph{Modern Visual Query Representations and Design trade-offs}:
	We use a few queries over intuitive database schemas 
	to discuss the main visual representations for relational queries 
	proposed by the database community.

Slides and videos of the tutorial will be made available afterwards on the tutorial web page,\footnote{\url{https://northeastern-datalab.github.io/visual-query-representation-tutorial/}} similar to other recent tutorials by the presenter and collaborators on unrelated topics.\footnote{\url{https://northeastern-datalab.github.io/topk-join-tutorial/}}\footnote{\url{https://northeastern-datalab.github.io/responsive-dbms-tutorial/}}

\section{Tutorial information}

\smallsection{Audience and prerequisite}
This 90~min tutorial targets
researchers and practitioners 
who desire an intuitive introduction
to the history of 
and approaches for
visual representations of relational queries and logical statements,
and desire to see major commonalities across past major designs of visual languages.
The tutorial is best followed by being familiar with Relational Algebra (RA), Relational Calculus (RC) and the safety conditions to make them equivalent in expressiveness.
However, the tutorial is self-contained and includes 
a concise short-paced introduction into overall characteristics of relational languages.

\smallsection{Scope of this tutorial}
This tutorial surveys visual formalisms for representing \emph{relational queries}. 
The focus is on relationally complete formalisms whose expressiveness is equivalent to Relational Algebra (RA) and Relational Calculus (RC) and non-recursive Datalog with stratified negation.
In order to guide the discussion, the tutorial discusses mapping the visual alphabets of visual formalisms 
to expressions of RA and RC. It thus starts with a quick overview of RA and RC and their connection to first-order logic.
It discusses various extensions to the relationally complete fragments 
(such as groupings and recursion) only at the end if time permits.

\smallsection{Out-of-scope}
The tutorial does not discuss domain-specific visualizations, such as those developed for geographic information systems, time-series, and
spatio-temporal 
data~\cite{DBLP:conf/ieeevast/CorrellG16,DBLP:conf/chi/ManninoA18,DBLP:journals/tvcg/LeeLSKKP20,DBLP:conf/chi/BattleFDBCG16}.
Neither does it discuss 
dynamic
interaction with queries or data~\cite{10.14778/2732240.2732247}.

\smallsection{Related other tutorials}
A tutorial at SIGMOD'19~ \cite{DBLP:conf/sigmod/TangWL19} (``Towards Democratizing Relational Data Visualizations'') focused on ways to visualize data and languages that allow users to specify what visualizations they want to apply to data. 
The focus of this tutorial is instead of visual representations \emph{of queries}.
Two tutorials at SIGMOD'17~\cite{DBLP:conf/sigmod/BhowmickCL17} (``Graph Querying Meets HCI'') and SIGMOD'22~\cite{DBLP:conf/sigmod/BhowmickC22} (``Data-driven Visual Query Interfaces for Graphs'') focused on visual composition of graph queries. 
The types of queries discussed in those tutorials basically correspond to conjunctive queries with inequalities over binary predicates, whereas our focus is on full-first order logic. 
Also the focus was on the human-interaction aspect of how to compose queries, while our focus is on the visual formalisms developed for relational queries over the last century (thus even predating the relational model).

\smallsection{Prior offerings of this tutorial} 
An early version of this tutorial was presented at the ``\emph{International Conference on the Theory and Application of Diagrams 2022}'' (DIAGRAMS-22)~\cite{gatterbauer:diagrams:tutorial:2022},
which is the main international venue covering all aspects of research on the theory an application of diagrams and attracts and audience complementary to the audience at VLDB. 
While that prior 
tutorial had a stronger focus 
on the 3rd (logical foundations of query languages) 
and 4th parts (diagrams predating relational databases), 
this tutorial will emphasize 
2nd (principles of query visualization) and
5th parts (visual query languages developed in the database community).

\section{Tutorial content}

\subsection{Why visualizing queries and why now?}

The tutorial starts by giving several scenarios 
in which ``appropriate'' query visualizations could help users 
achieve new functionalities 
or increased efficiency in composing queries.
An important detail is here that visualizations can be used as \emph{complement} 
to query composition,  
\emph{instead of substitution} for textual input.
This contrasts with 
Visual Query Languages (VQLs) which allow users to express queries in a visual format.
Visual methods for specifying relational queries have been studied
extensively~\cite{DBLP:journals/vlc/CatarciCLB97},
and
many commercial database products offer some visual interface for users to write simple conjunctive queries.  
In parallel, there is a centuries-old history on the study of formal diagrammatic reasoning systems \cite{DBLP:conf/iccs/Howse08}
with the goal of helping humans to reason in terms of logical statements.\footnote{A relational query is a logical formula with free variables. 
A logical statement has no free variables and is intuitively the same as a Boolean query that returns a truth value of TRUE or FALSE. }

Yet despite their  intuitive appeal and extensive study, successful visual tools today mostly only 
\emph{complement instead of replace}
text for composing queries.
We will discuss several reasons for why visual query composition
\emph{for general relational queries} 
have not yet widely replaced textual query composition
and discuss a user-query interaction 
that separates the query composition from the visualization:
Composition is unchanged and still done in text 
(or alternatively with exploratory input
formats like natural language).
But composition is augmented and \emph{complemented with a visual that helps interpretation}~\cite{gatterbauer2011databases}. 
With this motivation, the goal of this tutorial is to survey and highlight the key ideas behind
major proposals for diagrammatic representations of relational statements and queries.

\begin{definition}[Query Visualization~\cite{GatterbauerDJR:PrinciplesOfQV:2022}]
The term ``query visualization'' refers to both ($i$) a graphical representation of a query 
and ($ii$) the process of transforming a query into a graphical representation.
The goal of query visualization is to help users more quickly understand the intent of a query,
as well as its relational query pattern.
\end{definition}

\subsection{Principles of Query Visualization}

The challenge of query visualization is to find appropriate visual metaphors 
that ($i$) allow users to quickly understand a query's intent, even for complex queries,
($ii$) can be easily learned by users,
and ($iii$) can be obtained from textual queries by automatic translation, 
including a visually-appealing automatic arrangement of elements of the visualization.
We discuss 8 recently proposed principles of query visualization~\cite{GatterbauerDJR:PrinciplesOfQV:2022},
however newly organized, extended, and rephrased in the terminology of ``Algebraic Visualization Design''~\cite{DBLP:journals/tvcg/KindlmannS14} (\cref{Fig_QueryVisualizationAlgebraic}).
We refer to them later extensively when discussing different visualizations.
We also include those in order to spark a healthy debate during and after the tutorial.

\subsection{Logical foundations of relational 
languages}

We give a concise but comprehensive overview of \emph{the logical foundations of relational query languages}.
This overview uses a consistent notation that establishes shared concepts across relationally complete textual query languages.
We will reuse those extensively later when discussing visual query representations where we establish direct mappings between a given visual formalisms and logically equivalent textual queries. 
These mappings allow us a unified comparison of visual alphabets and their ``pattern expressiveness".
Thus our focus is on expressiveness basically equivalent to first-order logic, 
which allows us to connect a century of research on formalisms for diagrammatic reasoning to our topic.

\subsection{Early diagrammatic representations}

Relational calculus is a specialization of First-Order Logic (FOL), namely expressions with free variables. 
Diagrammatic representations for logical statements~\cite{DBLP:conf/iccs/Howse08} 
have been developed even before FOL, which was only clearly articulated in the 1928 first edition of David Hilbert and Wilhelm Ackermann's ``Grundzüge der theoretischen Logik''~\cite{HilbertAckerman:1928}.
An influential diagrammatic notation is the \emph{existential graph notation} by Charles Sanders Peirce~\cite{peirce:1933,Roberts:1992,Shin:2002}, 
who wrote on graphical logic as early as 1882~\cite{Peirce:vo4:1879-1884}.
These graphs exploit topological properties, such as enclosure, to represent logical expressions and set-theoretic relationships.
Peirce's graphs come in two variants: 
alpha and beta.
Alpha graphs represent propositional logic, whereas beta graphs represent first-order logic (FOL). 
Both variants use so-called \emph{cuts} to express negation (similar to our nesting boxes),
and beta graphs use a syntactical element called the \emph{Line of Identity} (LI) to denote 
\emph{both the existence of objects and the identity between objects}.
An important component of our discussions of Beta-existential graphs is showing their
imperfect mapping to the Boolean fragment of restricted forms of DRC.
As we will show, this imperfection has led to a lot of follow-up and confusions in various work on Peirce's existential graphs.
We may also shortly cover the close connections to 
Euler diagrams,
Venn diagramms, and
Venn-Peirce diagrams,
following the exposition by Shin~\cite{shin_1995}.

\subsection{Modern Visual Query Representations}

We discuss the main proposed visual representations for relational queries. 
We will also include influential Visual Query Languages (VQLs) as long as those support (either directly or via simple additions) the inverse functionality of visualizing an existing relational query.
A key difference of our tutorial in contrast to all prior surveys and overviews that we are aware of (like \cite{DBLP:journals/vlc/CatarciCLB97}) is 
that this tutorial shows original figures by using 
a consistent schema (the sailor-boat-database from the ``cow book'' \cite{RamakrishnanGehrke:DBMS2000}) 
and a few intuitive queries (such as ``find sailors who have rented all red boats'') 
to provide a consistent comparison across different past proposals.

\emph{Query-By-Example (QBE)}~\cite{DBLP:journals/ibmsj/Zloof77} is an influential early VQL that was strongly influenced by DRC.
QBE can express relational division 
breaking the query into two logical steps and using a temporary relation~\cite[Ch. 6.9]{RamakrishnanGehrke:DBMS2000}.
But in doing so, QBE
uses the query pattern from RA of implementing relational division (or universal quantification)
in a dataflow-type, sequential manner, 
requiring multiple occurrences of the same table.

\emph{Interactive query builders} employ 
visual diagrams that 
users can manipulate (most often in order to select tables and attributes)
while using \emph{a separate query configurator}
(similar to QBE's condition boxes~\cite{DBLP:journals/ibmsj/Zloof77}) 
to specify selection predicates, attributes, and sometimes nesting between queries.
They work mainly for constructing conjunctive queries
but limited forms of negation and union can be incorporated into the condition part of such queries. 
For more general forms of negation and union, however, views as intermediate relations need to be used, resulting in multiple screens.
dbForge~\cite{dbforge} is the most advanced and commercially supported tool we found for interactive query building.
Yet it does not show any visual indication for non-equi joins between tables 
and the actual filtering values and aggregation functions can only be added in a separate query configurator.
Moreover, it has limited support for nested queries: 
the inner and outer queries are built separately,
and the diagram for the inner query is \emph{presented separately and disjointly} 
from the diagram for the outer query.
Thus  \emph{no visual depiction of correlated subqueries is possible}.
Other graphical SQL editors like SQL Server Management Studio (SSMS)~\cite{ssms}, Active Query Builder~\cite{activequerybuilder}, QueryScope from SQLdep~\cite{queryscope}, MS Access \cite{msAccess}, and
PostgreSQL's pgAdmin3~\cite{pgadmin} lack in even more aspects of visual query representations: 
most do not allow nested queries, 
none has a single visual element for the logical quantifiers 
\texttt{NOT} \texttt{EXISTS} or \texttt{FOR} \texttt{ALL},
and all require specifying details of the query in SQL or across several tabbed views 
\emph{separate from a visual diagram}.

\emph{Dataflow Query Language (DFQL)} 
is an example visual representation that is relationally complete \cite{DBLP:journals/iam/ClarkW94,DBLP:journals/vlc/CatarciCLB97}
by mapping its visual symbols to the operators of relational algebra.
Following the same procedurality as RA, DFQL expresses the dataflow in a top-down tree-like structure. 
Like most visual formalisms
that we are aware of and that were proven to be relationally complete
(including \cite{DBLP:conf/sigmod/BakkeK16} and those listed in
\cite{DBLP:journals/vlc/CatarciCLB97})
they are at their core visualizations of relational algebra operators.
This applies even to the more abstract \emph{graph data structures (GDS)} from \cite{DBLP:conf/vdb/Catarci91} and the later \emph{graph model (GM)} from \cite{DBLP:journals/is/CatarciSA93}.
The key difference is that GDS and GM are formulated inductively based on mappings onto operators of relational algebra. 
They thus mirror dataflow-type languages where visual symbols (directed hyperedges) represent operators like \emph{set difference} connecting two relational symbols, 
leading to a new third symbol as output.

\emph{DataPlay}~\cite{DBLP:conf/uist/AbouziedHS12} 
uses a nested universal relation data model and
allows a user to compose their query by interactively modifying a \emph{query tree with quantifiers}
and observing changes in the matching/non-matching data.
\emph{Visual SQL}~\cite{DBLP:conf/er/JaakkolaT03} is 
a visual query language that also support query visualization. 
With its focus on query specification, it maintains the one-to-one correspondence to SQL,
and syntactic variants of the same query lead to different representations.
Similarly, \emph{SQLVis}~\cite{DBLP:conf/vl/MiedemaF21} 
places a strong focus on the actual syntax of a SQL query 
and syntactic variants like nested EXISTS queries change the visualization.
\emph{GraphSQL}~\cite{DBLP:conf/dexaw/CerulloP07} uses visual metaphors that are different from typical relational schema notations 
and visualizations, even simple conjunctive queries can look unfamiliar.
The \emph{Query Graph Model (QGM)} developed for Starburst~\cite{DBLP:conf/sigmod/HaasFLP89}
helps users understand query plans, not query intent.
QueryVis (earlier QueryViz)~\cite{DanaparamitaG2011:QueryViz, 
DBLP:journals/tvcg/BartolomeoRGD22,
gatterbauer2011databases,
DBLP:conf/sigmod/LeventidisZDGJR20}
borrows the idea of a ``default reading order''
from diagrammatic reasoning systems~\cite{DBLP:conf/diagrams/FishH04} 
and uses \emph{arrows} to indicate an implicit reading order between different nesting levels.
Without the arrows, there would be no natural order placed on the existential quantifiers 
and the visualization would be ambiguous.
\queryvis focuses on the non-disjunctive fragment of relational calculus and is guaranteed to represent connected nested queries unambiguously up to nesting level 3.
\diagrams~\cite{relationalDiagrams}
is a more recent variant that indicates the nesting structure of table variables by using \emph{nested negated bounding boxes} (instead of arrows)
inspired by Peirce's influence beta existential graphs~\cite{peirce:1933,Roberts:1992,Shin:2002}.
Interestingly, because \diagrams are based on Tuple Relational Calculus (instead of Domain Relational Calculus which is closer to First-Order Logic)
they solve interpretation problems of beta graphs that have been the focus of intense research in the diagrammatic reasoning communities.

\section{Author information}

\textbf{Wolfgang Gatterbauer} is 
an Associate Professor at the Khoury College of Computer Sciences at Northeastern University.
His research interests lie in the intersection of theory and practice of data management.
With co-authors he got the EDBT 2021 best paper award, ``best of conference'' mentions for 
PODS 2021, SIGMOD 2017, WALCOM 2017, and VLDB 2015, and two SIGMOD 2021 reproducibility awards. 
Prior to joining Northeastern, he was an Assistant Professor at Carnegie Mellon's Tepper School of Business,
and before that a PostDoc at University of Washington.
He received his PhD in Computer Science at Vienna University of Technology.

\begin{acks}
This work was supported in part by NSF under award numbers IIS-1762268 and IIS-1956096.
\end{acks}

\bibliographystyle{ACM-Reference-Format}
\balance
\bibliography{queryvis-vldb-tutorial.bib}

\end{document}